   \documentclass[prl,aps,twocolumn,showpacs]{revtex4}
\usepackage[dvips]{graphicx}
\usepackage{dcolumn}%
\usepackage{amsmath}%
\usepackage{amsfonts}%
\usepackage{amssymb}
 \usepackage[usenames]{color}
\providecommand{\U}[1]{\protect\rule{.1in}{.1in}}
\newcommand{\be}{\begin{equation}}
\newcommand{\ee}{\end{equation}}
\newcommand{\bea}{\begin{eqnarray}}
\newcommand{\eea}{\end{eqnarray}}
\newcommand{\bt} {\begin{tabular}}
\newcommand{\et} {\end{tabular}}
\newcommand{\nn}{ \nonumber}

\newcommand{\ba} {\begin{align}}
\newcommand{\ea} {\end{align}}
\begin{document}

\title{Nonlinear thermoelectric transport in single-molecule junctions: the effect of electron-phonon interactions}

\author{Natalya A. Zimbovskaya}

\affiliation
{Department of Physics and Electronics, University of Puerto 
Rico-Humacao, CUH Station, Humacao, Puerto Rico 00791, USA}

\begin{abstract}
  In the present work, we theoretically analyze the steady-state thermoelectric transport through a single-molecule junction with a vibrating bridge. Thermally induced charge current in the system is explored using a nonequilibrium Green's functions formalism. We study combined effects of Coulomb interactions between charge carriers on the bridge and electron-phonon interactions on the thermocurrent beyond the linear response regime. It is shown that electron-vibron interactions may significantly affect both magnitude and direction of the thermocurrent and vibrational signatures may appear.
   \end{abstract}


\date{\today}
\maketitle

\section{i. introduction}

Presently,  thermoelectric transport in nanoscale systems attracts much interest which was triggered in early nineties by pioneering works of Hicks and Dresselhaus \cite{1,2}. They predicted that heat-to-electricity conversion efficiency known to be rather low in conventional bulk materials may be significantly enhanced in nanoscale systems. Correspondingly, thermoelectric properties of tailored nanoscale systems such as carbon-based nanostructures and quantum dots (QD) and/or molecules put in contact with macroscopic leads were explored both theoretically and experimentally \cite{3,4,5,6}. 

The key to transport properties of single-molecule junctions is a combination of discrete electron energy spectrum of a molecular bridge and nearly continuous energy spectra typical for charge carriers on the leads. Owing to this combination, sharp features appear in the electron transmission spectra which determine various transport characteristics. In particular, these features may give rise to a considerable enhancement of the heat-to-electricity conversion efficiency in single-molecule junctions and  similar nanoscale systems \cite{5,6}.

Transport through molecular junctions is controlled by several factors including the strength of the molecular bridge coupling to the leads and Coulomb interactions between traveling electrons. Also, interactions between electrons and molecular vibrations  may strongly affect electron transport through molecules. Over the past two decades, theoretical studies of vibrationally-induced electron transport through QDs and single-molecule junctions was carried out by many authors. Mostly, they used master equations \cite{7,8,9,10,11}, functional renormalization group based approaches \cite{12,13}, scattering theory \cite{14,15,16,17}, and nonequilibrium Green's functions formalism (NEGF) \cite{18,19,20,21,22,23,24,25,26}.  Signatures of electron-vibron interactions were observed in experiments \cite{27,28,29,30,31}.

In the present work we focus on Seebeck effect in molecular junctions. As known, a difference in the leads temperatures induces a charge current $I_{th} $ flowing through the system. Seebeck effect is measured by recording the voltage $ V_{th} $ which completely stops this current provided that the difference in temperatures $ \Delta T $ remains fixed. When $ \Delta T \ll T_{L,R}\ (T_{L,R} $  being the temperatures of the left and right electrode, respectively), the system operates within a linear response regime, and both thermally excited current $ I_{th} $ and thermovoltage $ V_{th} $ are proportional to $ \Delta T. $ However, as the temperature gradient across the system increases, the system may switch to nonlinear regime of operation. Nonlinear Seebeck effect was observed in experiments on semiconducting QDs, single-molecule junctions \cite{32,33}, and in magnetic tunnel junctions \cite{34}. Therefore, the discussion of Seebeck effect in nanoscale systems was extended to the nonlinear regime \cite{8,25,26,34,35,36,37,38,39}. We remark that properties of thermovoltage in nanoscale systems were studied more thoroughly than those of thermally excited current in spite of the fact that $ I_{th} $ is more convenient for measuring and modeling. In the present work, we contribute to studies of Seebeck effect in nanoscale systems by analyzing the effect of electron-electron and electron-phonon interactions of thermally excited charge current flowing through a molecular junction beyond the linear in $ \Delta T $ regime. 

\section{ii. Model and main equations}

Present calculations are based on the Anderson-Holstein model for a molecular junction. We simulate the molecule linking the leads by a single energy level and a single vibrational mode. A schematic drawing of the considered system is shown in the Fig. 1. The mode is linearly coupled to electrons on the molecular bridge as well as to a phonon bath. This model is commonly used to theoretically analyze diverse aspects of electron transport through molecular junctions and other similar systems.The relevant Hamiltonian may be written as  $ H = H_D + H_L + H_R + H_T + H_{ph}. $ Here, $ H_D $ represents the single-level molecular bridge coupled to the vibrational mode: 
\begin{align}
H_D = & \sum_\sigma E_\sigma d_\sigma^+ d_\sigma + U d_\sigma^+ d_\sigma d_{-\sigma}^+ d_{-\sigma}  + \hbar\Omega a^+ a 
\nn\\ &
+ \Lambda Q_a \sum_\sigma c_\sigma^+ c_\sigma   \label{1}
\end{align}
In this expression, $ d_\sigma^+\ (d_\sigma) $ creates (annihilates) an electron with spin $\sigma $ on the bridge, $ E_\sigma = E_0 $ is the energy of a single spin-degenerated bridge level, $ U $ is the charging energy, $ a^+(a^-) $ creates (degenerates) the vibrational mode with the frequency $ \Omega, $ and $ Q_a = a^+ + a^-. $ The last term describes electron-vibron interactions with $ \Lambda $ being the relevant coupling parameter. The terms $ H_\beta\ (\beta = L,R) $ are corresponding to noninteracting electrons on the leads with energies $ \epsilon_{r\beta\sigma}: $
\be
H_\beta = \sum_{r\sigma} \epsilon_{r\beta\sigma} c_{r\beta\sigma}^+ c_{r\beta\sigma}      \label{2}
\ee
where $ c_{r\beta\sigma}^+ $ and $ c_{r\beta\sigma} $ are creation and annihilation operators for these electrons. The transport term:
\be 
H_T = \sum_{r\beta\sigma} \tau_{r\beta\sigma} c_{r\beta\sigma}^+ d_\sigma + H.C \label{3}
\ee
represents tunneling effects between the bridge and the leads, factors $ \tau_{r\beta\sigma} $ characterizing the coupling of electron states on the bridge to those on the leads. Below we consider a symmetrically coupled system, so $ \tau_{rL\sigma} = \tau_{rR\sigma} = \tau_{r\sigma}. $ Finally, the term $ H_{ph} $ describes the phonon bath coupled to the vibrational mode. We assume that the bath is kept at the temperature $ T = \frac{1}{2}(T_L + T_R), $ and that its coupling to the vibrational mode far exceeds the electron-phonon coupling strength. Then the phonon population on the bridge is given by Bose-Einstein distribution $ N_{ph} $ \cite{24}.  

To eliminate the electron-phonon coupling term from the Hamiltonian (\ref{1}), a small polaron transformation is commonly
applied to convert the Hamiltonian  $ H $ as $ \tilde H = \exp (s) H\exp (-s) $ where 
\be
s = \frac{\Lambda}{\hbar\Omega} \sum_\sigma d_\sigma^+ d_\sigma (a^+ - a) .  \label{4}
\ee
As a result, the term describing electron-phonon interactions disappears from the transformed bridge Hamiltonian (\ref{1}), and the energies $ E_0 $ and $U $ acquire corrections originating from electron-phonon interactions and should be replaced by renormalized energies:
\be
\tilde E_0 = E - \frac{\Lambda^2}{\hbar\Omega},\qquad
\tilde U = U - \frac{2\Lambda^2}{\hbar\Omega}.  \label{5}
\ee
The transfer Hamiltonian (\ref{3}) also undergoes a transformation, so that renormalized coupling parameters $ \tilde\tau_{r\sigma} $ are substituted for $ \tau_{r\sigma} :$
\be
\tilde\tau_{r\sigma} = \tau_{r\sigma}X \equiv \tau_{r\sigma} \exp \left[-\frac{\Lambda}{\hbar\Omega} (a^+ - a) \right].
\label{6} \ee
The expectation value of the operator $ X $ in the considered case of thermal equilibrium is given by:
\be
\big<X\big> = \exp \left[-\left(\frac{\Lambda}{\hbar\Omega}\right)^2 \left(N_{ph} + \frac{1}{2}\right) \right].
\label{7}  \ee
Following Ref. \cite{21}, one may approximate the expression for $ \tilde\tau_{r\sigma} $ by substituting $ \big<X \big> $ instead of $ X $ into Eq. (\ref{6}),  thus decoupling the electron and phonon subsystems.

The transformed Hamiltonian $ \tilde H $ is the sum of the electron part $ \tilde H_{el} $ and the phonon part  $ \tilde H_{ph}.$ Since the electron and phonon subsystems are decoupled, these Hamiltonians may be applied to independently compute relevant average values. The Hamiltonian $ \tilde H_{el} $ has the form similar to that describing a single-level bridge attached to the leads via transfer terms. However, the key parameters of the model $ E_0, \ U $ and $ \tau_{r\sigma} $ are now replaced by renormalized values given by Eqs. (\ref{5})-(\ref{7}). Using this Hamiltonian, one can derive expressions for renormalized electron Green's functions. Disregarding spin-flip processes, one arrives at separate equations for the retarded and advanced Green's functions corresponding to different spin orientations. The expressions for the retarded Green's function can be presented in the form first derived in Ref. \cite{40}:
\be
\tilde G_\sigma^{rr} (E) = \frac{E - \tilde E_0 - \tilde \Sigma_{02}^\sigma - \tilde U(1 - \big<n_{-\sigma}\big>)}{(E - \tilde E_0 - \tilde \Sigma_{0\sigma})(E - \tilde E_0 - \tilde U - \tilde \Sigma_{02}^\sigma) + \tilde U \tilde\Sigma_{1\sigma}}.
\label{8}  \ee
Here, $ \big<n_{\sigma}\big> $ are one-particle occupation numbers on the bridge level:
\be
\big<n_{\sigma}\big> = \int \frac{dE}{2\pi} \mbox{Im} \big [\tilde G_\sigma^<(E)\big]
\label{9}   \ee  
and $\tilde \Sigma_{0\sigma},\ \tilde\Sigma_{1\sigma},$ and $\tilde\Sigma_{02} $ are self-energy corrections. These self-energy terms are described by usual expressions (see e.g. Refs. \cite{40,41,42}) where the characteristic energies $ E_0,\ U $ and $\tau_{r\sigma} $ are replaced by the renormalized values.

The lesser Green's function $ \tilde G^<(E) $ is related to the retarded and advanced Green's functions by Keldysh equation:
\be
\tilde G_\sigma^< (E) = \tilde G_\sigma^{rr}(E) \tilde \Sigma_\sigma^<(E) \tilde G_\sigma^{aa}(E)
\label{10}   \ee
A similar expression may be written out for the greater Green's function $\tilde G^>(E) $ by substituting a self-energy $ \tilde \Sigma^>(E) $ for $ \tilde \Sigma^<(E). $ These self-energy terms may be presented in the form \cite{41}:
\begin{align} &
\tilde \Sigma_\sigma^<(E) = i \sum_\beta f_\sigma^\beta (E) \Delta_\sigma^\beta(E), \label{11}
\\ &
\tilde \Sigma_\sigma^>(E) = - i \sum_\beta \big[1 - f_\sigma^\beta(E)\big] \Delta_\sigma^\beta(E).
\label{12}        \end{align}
Here, $ f_\sigma^\beta(E) $ are Fermi distribution functions for the leads, and factors $ \Delta_\sigma^\beta(E) $ differ from $\tilde \Gamma_\sigma^\beta(E) \equiv  - 2\mbox{Im} \big[\tilde\Sigma_{0\sigma}^\beta(E)\big] $ due to renormalization originating from Coulomb interactions between electrons on the bridge. When these interactions are omitted from consideration, the difference between $ \Delta_\sigma^\beta (E) $ and $ \tilde\Gamma_\sigma^\beta(E) $ vanishes. The terms $ \tilde \Gamma_\sigma^\beta(E) $ describe coupling of the bridge to the leads. Renormalization of these terms due to Coulomb interactions may bring noticeable changes into the values of occupation numbers when the considered system is shifted from the equilibrium position. In the present work, we assume that this occurs due  to the temperature gradient applied across the system. It was shown \cite{39} that for moderate values of $ \Delta T\ (\Delta T/T < 1) $ one may disregard the renormalization of coupling parameters for it brings very small corrections to $ \big< n_\sigma \big> $ values. Therefore, in further calculations we employ Eqs. (\ref{11}), (\ref{12})  where $  \Delta_\sigma^\beta(E) $ are replaced by  $ \tilde \Gamma_\sigma^\beta(E) ,$
and the latter parameters are computed using the wide band approximation.

True Green's functions for the electrons on the bridge are related to those given before. For instance, one may approximate the lesser and greater Green's functions as follows \cite{21}:
\begin{align}
G_\sigma^<(E) = \sum_{r =-\infty}^\infty L_r \tilde G_\sigma^<(E + r\hbar\Omega), \label{13}
\\
 G_\sigma^<(E) = \sum_{r =-\infty}^\infty L_r \tilde G_\sigma^<(E - r\hbar\Omega),  
\label{14}   \end{align}
with the coefficients $ L_r $ of the form:
\begin{align}
L_r = & \exp \left[ - \left(\frac{\Lambda}{\hbar\Omega}\right)^2 (2N_{ph} + 1) + \frac{r \hbar\Omega}{2kT} \right] 
\nn\\   & \times
I_r \left[\frac{2\Lambda}{\hbar\Omega} \sqrt{N_{ph} (N_{ph} + 1 )} \right].
\label{15}    \end{align}
Here, the Bose-Einstein distribution function $ N_{ph} $ is taken for the phonon frequency $ \Omega $ and the temperature $ T, \ k $ is the Boltzmann's constant and $ I_r (z) $ is the modified Bessel function of the order $ "r". $  

Assuming that $f_\sigma^\beta(E) \equiv f^\beta(E) $ we may present the charge current flowing through a symmetrically coupled junctions as the Landauer expression:
\be
I = \frac{e}{\pi\hbar} \int \tau(E) \big[f^L(E) - f^R (E)\big] dE
\label{16}    \ee
where the electron transmission function is given by:
\be
\tau (E) = \frac{i}{4} \Gamma \sum_\sigma \left[G_{\sigma\sigma}^< (E) - G_{\sigma\sigma}^> (E) \right].
\label{17}    \ee

We remark that in this expression, the prefactor  $\Gamma $ characterizing the coupling of the leads to  the bridge $(\Gamma_\sigma^L = \Gamma_\sigma^R \equiv \Gamma)$ is computed omitting renormalization due to electron-phonon interactions.
Strictly speaking, Eqs. (\ref{16}), (\ref{17}) remain valid only when electron-phonon interactions are sufficiently weak $ (\Lambda \lesssim \hbar\Omega). $ This conclusion is resulting from the procedure used to obtain the expressions (\ref{13})-(\ref{15}). In deriving these expressions, only terms of the lowest order in $ (\Lambda/\hbar\Omega) $ were taken into account. Better approximations for the relevant Green's functions were obtained is several works (see e.g. Refs. \cite{20,26}). However, in these works, the effects of Coulomb interactions between electrons on the bridge were not considered. Keeping in mind the above mentioned limitations, we employ Eqs. (\ref{13})-(\ref{17}) to study steady-state thermoelectric transport through a molecular junction with the vibrating bridge.

\section{iii. Numerical results}

In this section, we consider some numerical results obtained by applying the outlined above formalism. The effect of electron-vibron interactions on transport characteristics of a molecular junction strongly depends on the relationships between the energies $ \Gamma,\hbar\Omega $ and $ \Lambda. $ It was argued that when the bridge is weakly coupled to the leads $ (\Gamma \ll \hbar\Omega, \Lambda) $ the "mobility" of traveling electrons remains low and the local oscillator (the vibrational mode) is fast enough to individually adjust to them \cite{4,26}. This creates the most favorable conditions for manifestations of electron-vibron interactions in transport characteristics of single-molecule junctions, so we assume this condition to be satisfied. Also, we assume that  electron-phonon interactions are moderate  $(\Lambda < \hbar\Omega)$ and electron-electron interactions are sufficiently strong $(U > \hbar\Omega) .$ Reasonable  values for these energies may be taken basing on the fact that frequencies of vibrational modes range between $ 10 meV $ and several hundreds of $ meV $ \cite{28,31}. In further calculations we accept $ \hbar\Omega = 10 meV $ which is typical for small molecules.  

\begin{figure}[t] 
\begin{center}
 \includegraphics[width=3.2cm,height=4cm,angle=-90]{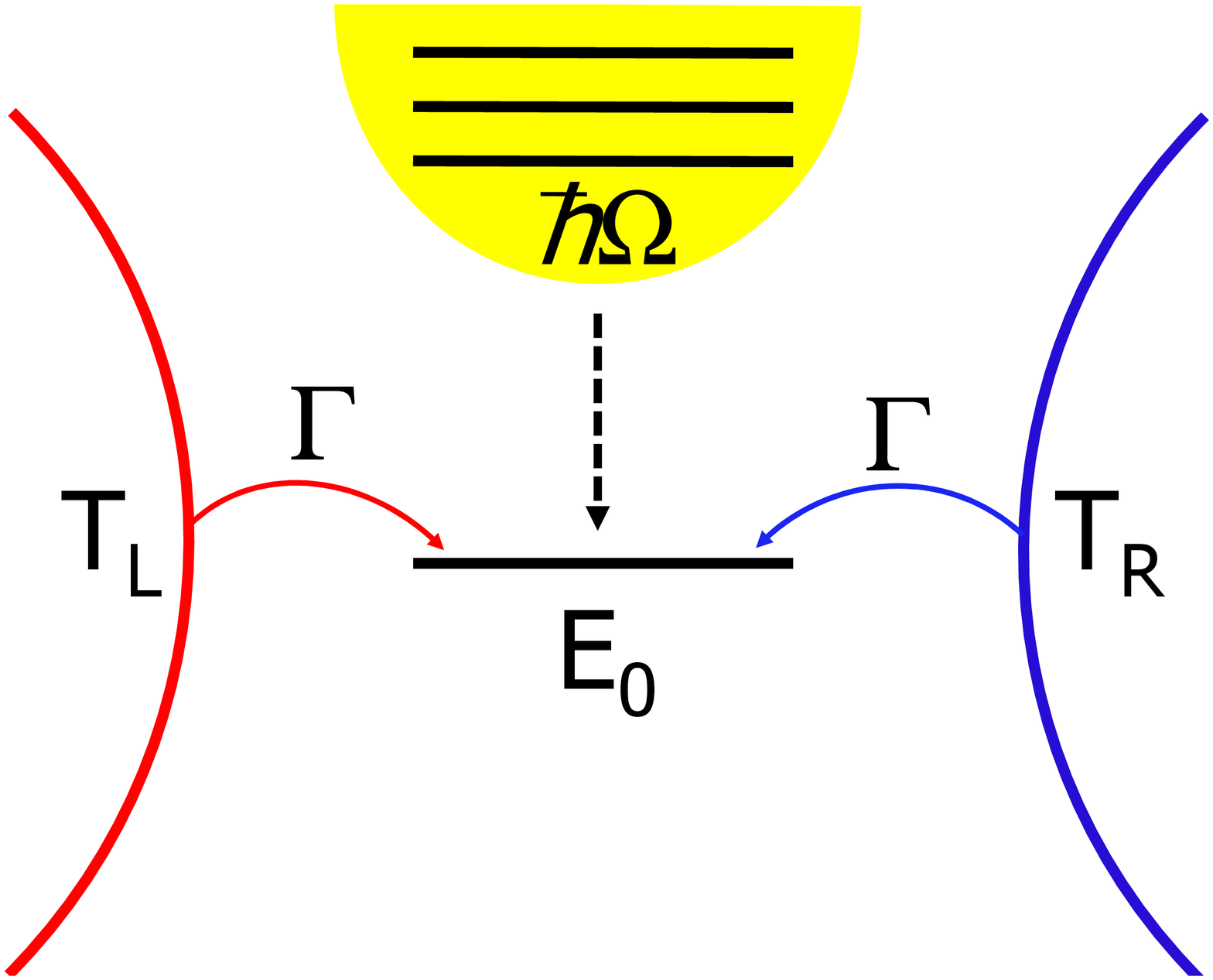}
\includegraphics[width=8.8cm,height=4.4cm]{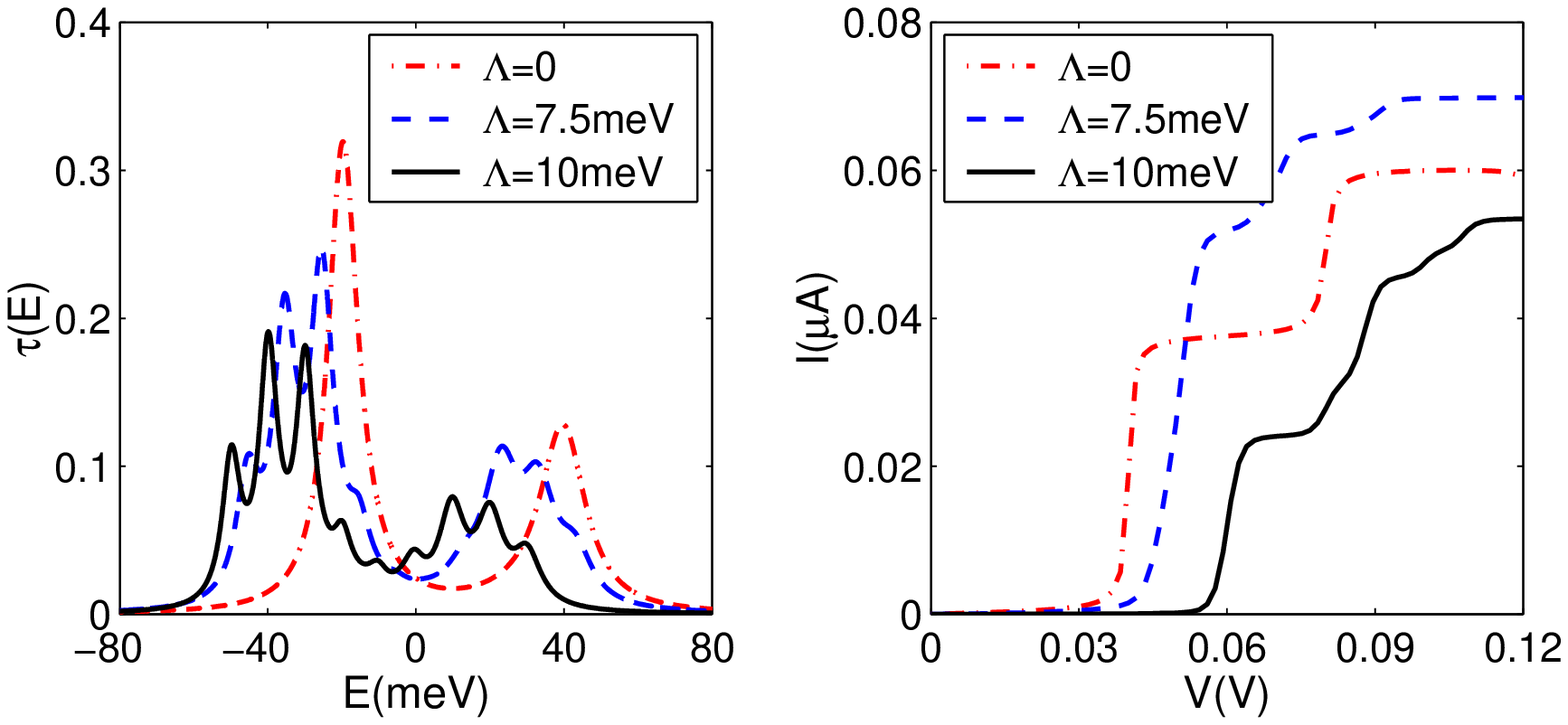} 
\caption{(Color online)  Top panel:  A schematic drawing of the considered junction where electrodes are kept at different temperatures $ T_L $ and $ T_R\ (T_L > T_R) $ and a single-level bridge is coupled to the vibrational mode with the frequency $ \Omega. $ 
Bottom panels: The electron transmission through a junction with a single-level bridge (left) and the bias voltage induced current flowing through the junction (right) affected by electron-phonon interactions. The curves are plotted for $ kT_L = kT_R = 0.6meV,\ U = 60meV,\ \hbar\Omega = 10meV,\ \Gamma = 2.5meV $ (left panel) and $ \Gamma = 0.25 $ (right panel). 
}
 \label{rateI}
\end{center}\end{figure}

While on the bridge, electrons participate in events occurring due to their interactions with vibrational phonons. These events involve virtual phonon emission and absorption resulting in appearance of metastable electron levels. within the adopted model where the bridge is simulated by a single level coupled to a single vibrational mode, these states have energies close to $ E_n =\tilde E_0 + n\hbar\Omega  $ and  $ E_m = \tilde E_0 + \tilde U + m \hbar\Omega\ (n,m = 0,1,2, \dots). $ At weak coupling of the bridge to the leads, these states have sufficiently long lifetimes, so they may serve as extra channels for electron transport. Signatures of these states may appear in the electron  transmission, as shown in the Fig. 1. In the presence of electron-vibron interactions, each of the two unequal in height peaks corresponding to electron transmission through a junction with a single-state bridge within the Coulomb blockade regime is replaced by a set of narrower peaks associated with the metastable states. All these peaks are arranged as two subsets including the peaks centered at energies $ E = \tilde E_0 $ and $ E =\tilde E_0 + \tilde  U ,$ respectively. 
One observes that the subsets become closer to each other as the electron-phonon coupling strengthens. This happens because the effective charging energy $ \tilde U $ is renormalized due  to electron-phonon interactions. As follows from Eqs. (\ref{6}), (\ref{7}), electron-phonon interactions significantly reduce the effective coupling of the bridge to the leads. Therefore, one may expect the considered junction to remain within the Coulomb blockade regime even at low temperatures. Nevertheless, it was shown that phonon-induced sidebands may accompany Kondo maximum in the electron transmission \cite{12,21}. Phonon-induced peaks in the transmission may give rise to extra steps in current-voltage curves provided that the bridge is extremely weakly coupled to the leads and the temperature is sufficiently low. These features are displayed in the right bottom panel of the Fig. 1.

\begin{figure}[t] 
\begin{center}
\includegraphics[width=8.8cm,height=4.4cm]{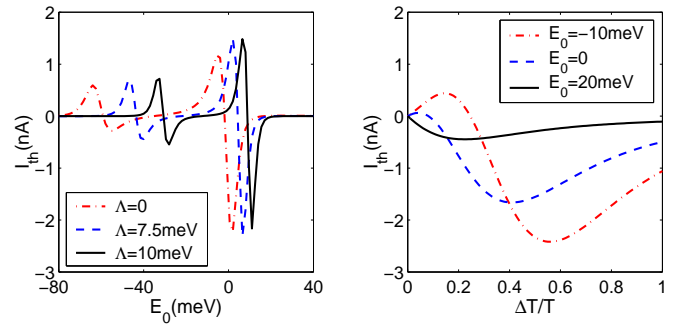} 
\caption{(Color online)  Thermally excited current $ I_{th} $ in an unbiased junction as a function of the bridge level energy $ E_0 $ at fixed value of $ \Delta T $ (left) and as a function of $ \Delta T $ at several values of $ E_0 $ (right).
The curves are plotted for $  kT_R = 0.6meV,\ U = 60meV,\ \hbar\Omega = 10meV,\ \Gamma = 2.5meV,\ \Delta T = 0.8 T_R $ (left panel) and $ \Lambda = 7.5meV $ (right panel). 
}
 \label{rateI}
\end{center}\end{figure}

 Thermally excited current starts to flow through an unbiased molecular junction when the leads are kept at different temperatures. For certainty, we assume below that the right lead is cooler than the left one $(T_R < T_L).$ Also, we assume that $ T_R $ remains constant whereas $ T_L $ varies. The difference of the Fermi distributions in the expression (\ref{16}) for the thermally excited   charge current takes on nonzero values solely when the tunnel energy values belong to a close vicinity of the chemical potential of the leads $ \mu. $  For simplicity, we assume that $ \mu = 0, $ thus the thermally excited current $ I_{th} $ flows through the system provided that the energy level on the bridge is shifted to a position where the renormalized energy $ \tilde E_0 $ (or $\tilde E_0 + \tilde U)$ is close to zero. This corresponds to opening transport channels for charge carriers. As $ \tilde E_0 $ is approaching zero from above, electrons start to flow from the left (hot) electrode to the right (cold) one, and $ I_{th} $ accepts negative values, as shown in the Fig. 2. However, when $ \tilde E_0 $ becomes very close to zero, holes start to participate in transport along with electrons. At certain value of $ \tilde E_0 $ the holes flow counterbalances the electron flow, and $ I_{th} $ becomes zero. At further shift of $ \tilde E_0 $ to a position below zero, the holes flow predominates, and $ I_{th} $ accepts positive values. 
  The same explanation could be given for $ I_{th} $ behavior at $ \tilde E_0 $ close to $ - \tilde U. $ 

Interactions of charge carriers on the bridge with the vibrational mode bring noticeable changes into $ I_{th} $ behavior. Renormalization of the bridge level energy $ E_0 $ and the charging energy $ U $ due to these interactions changes positions of the derivative-like features corresponding to opening up channels for charge carriers transport. One observes that the stronger electron-vibron interactions become, the smaller is the separation between these features. Also, $ I_{th} $ magnitude largely depends on the electron-vibron interaction strength. However, at fixed $ \Lambda,$ peaks in electron transmission corresponding to metastable states $ E_n $ and $ E_m \ (n,m \neq 0) $ do not leave explicit signatures in the $ I_{th}(E_0) $ lineshape shown in the figure. This agrees with the results reported in earlier works \cite{18,20,26}.

\begin{figure}[t] 
\begin{center}
\includegraphics[width=8.8cm,height=7.8cm]{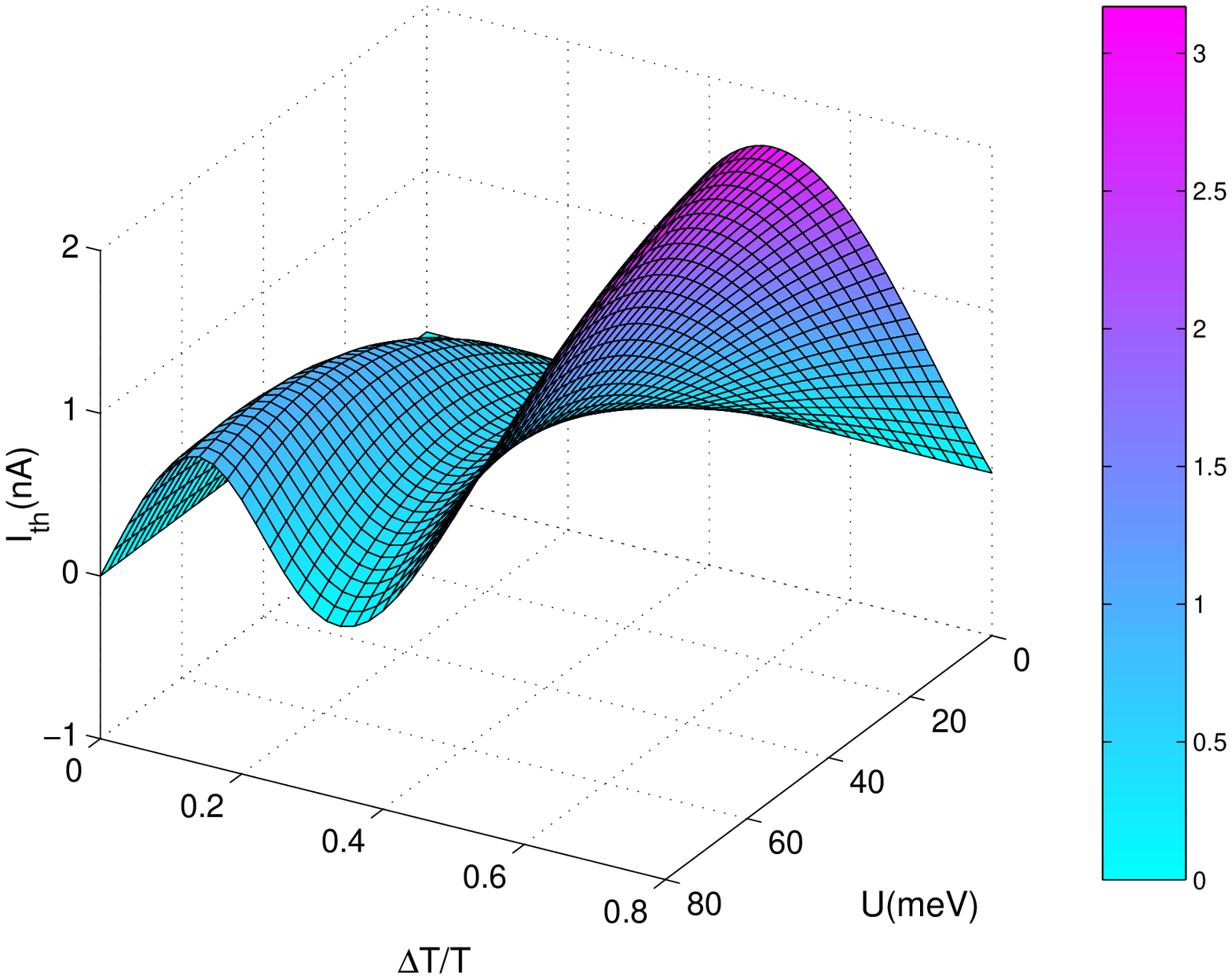} 
\includegraphics[width=8.8cm,height=4.5cm]{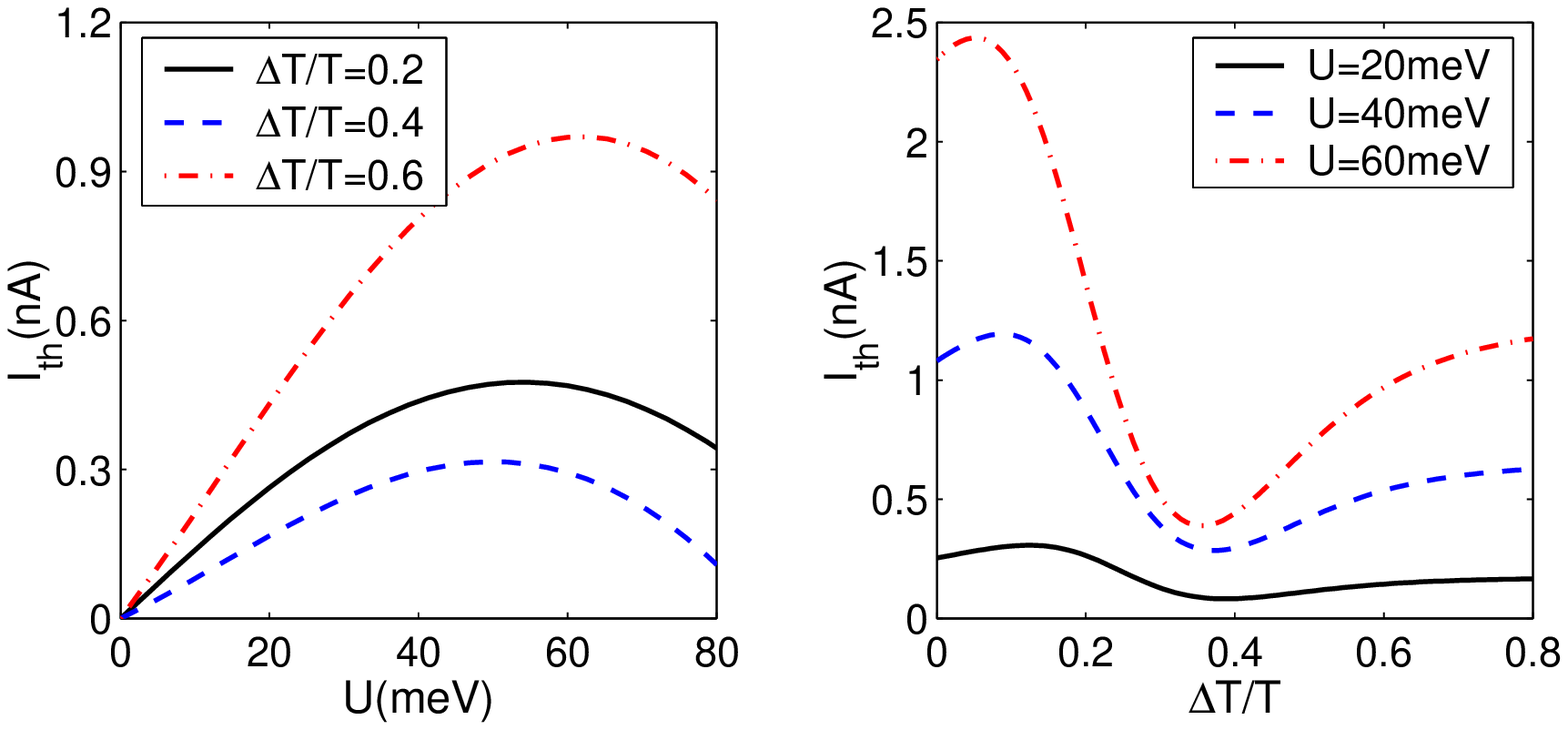} 
\caption{(Color online)  Top panel: Thermally excited current $ I_{th} $ versus $ \Delta T $ and $ U $ plotted assuming $ kT_R = 0.6meV,\ E_0 = -10meV,\ \Lambda = 7.5meV,\ \Gamma = 2.5meV,\  \hbar\Omega = 10meV.$ Bottom panels: Cross-sections of the surface shown on the top at several fixed values of $T_L $ (left) and at several fixed values of $ U$ (right).
}
 \label{rateI}
\end{center}\end{figure}

Thermally excited current strongly depends on the temperature difference $ \Delta T. $ As demonstrated in the Fig. 2, $ I_{th} $ accepts a distinctly nonlinear lineshape when $ \Delta T $ is not too small. To a considerable degree, $ I_{th} $ temperature dependence is controlled by the bridge energy level position. For each considered  value of $ E_0,\ I_{th}(\Delta T) $ reaches its minimum at certain value of $ \Delta T. $ Besides, some of these curves show maxima at small values of $ \Delta T $ followed by the change of sign.  As discussed if Refs. \cite{38,39}, the variety of $ I_{th} (\Delta T)$ lineshapes partly originates from the relationship between characteristic energies $ E_0 $ and $ U. $ The third controlling factor is the electron-vibron coupling strength $ \Lambda. $

\begin{figure}[t] 
\begin{center}
\includegraphics[width=8.8cm,height=7.8cm]{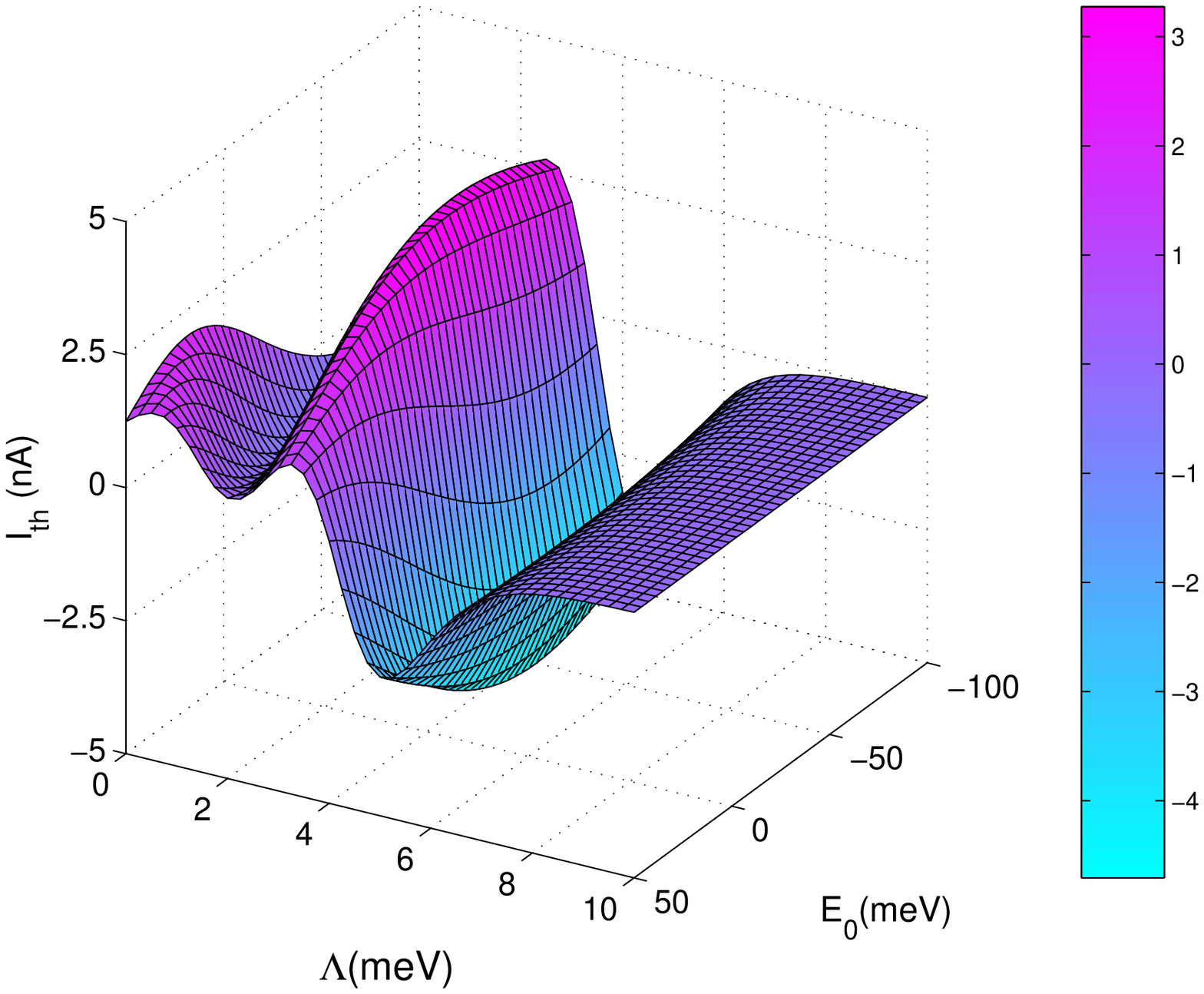} \includegraphics[width=8.8cm,height=4.5cm]{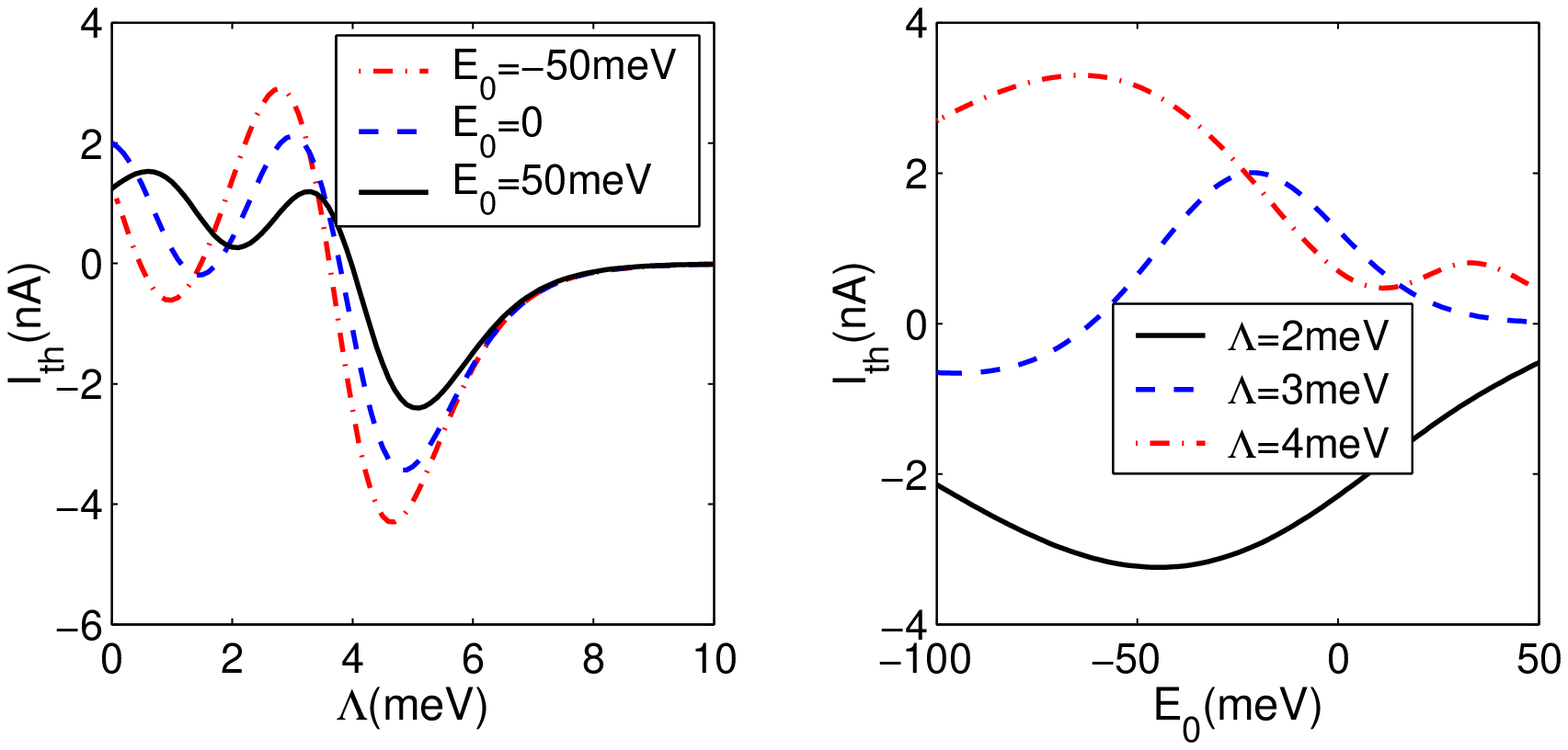} 
\caption{(Color online)  Top panel: Thermally excited current $ I_{th} $ versus $ E_0 $ and $ \Lambda $ plotted assuming $ kT_R = 6meV,\ E_0 = -10meV,\ \Delta T = 0.8 T_R,\ \Gamma = 2.5meV,\  \hbar\Omega = 10 meV.$ Bottom panels: Cross-sections of the surface shown on the top at several fixed values of $ E_0 $ (left) and at several fixed values of $ \Lambda $ (right).
}
 \label{rateI}
\end{center}\end{figure}

The effect of Coulomb interactions is further elucidated in Fig. 3. One observes that the value of the charging energy $ U $ affects positions of minima at $ I_{th} (\Delta T) $ curves as well as the thermocurrent magnitudes. The effect of electron-phonon coupling at comparatively high temperatures is illustrated in Fig. 4. The presented surface reveals several ridges separated by ravines which means that electron-vibron coupling may influence $ I_{th} $ in more than one way. This may be observed by studying this surface profiles corresponding to several fixed  $ E_0 $ values. A rather complex shape of $ I_{th} (E_0,\Lambda) $ surface may be explained as follows. It was demonstrated before that $ I_{th} $ takes on nonzero values when the renormalized energies $ \tilde E_0 $ and $ \tilde E_0 + \tilde  U $ are close to the chemical potential of electrodes. The width of "conduction window" centred at $ E = \mu = 0 $ is determined by the temperature difference $ \Delta T. $

\begin{figure}[t] 
\begin{center}
\includegraphics[width=8.8cm,height=7.8cm]{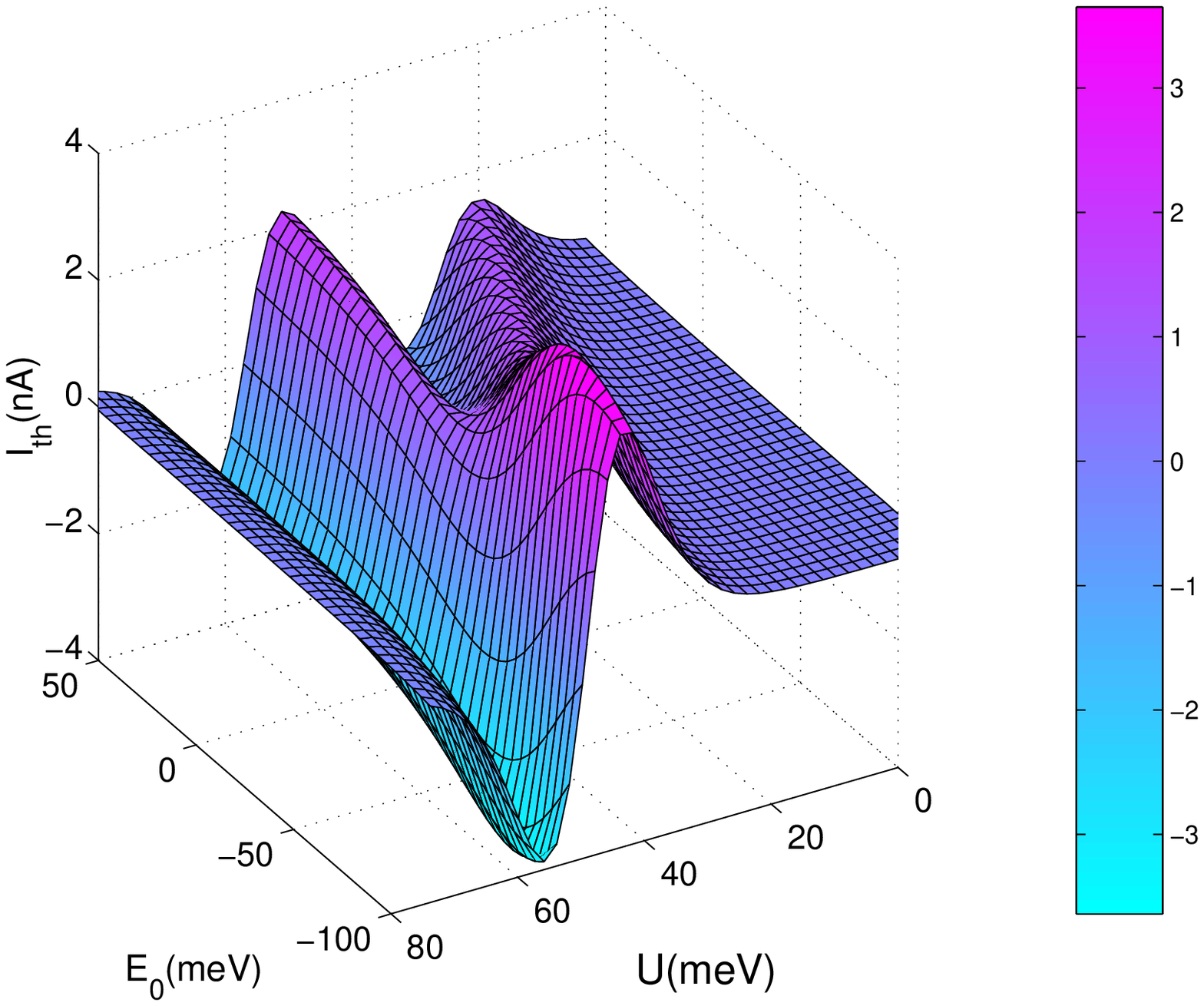} 
\includegraphics[width=8.8cm,height=4.5cm]{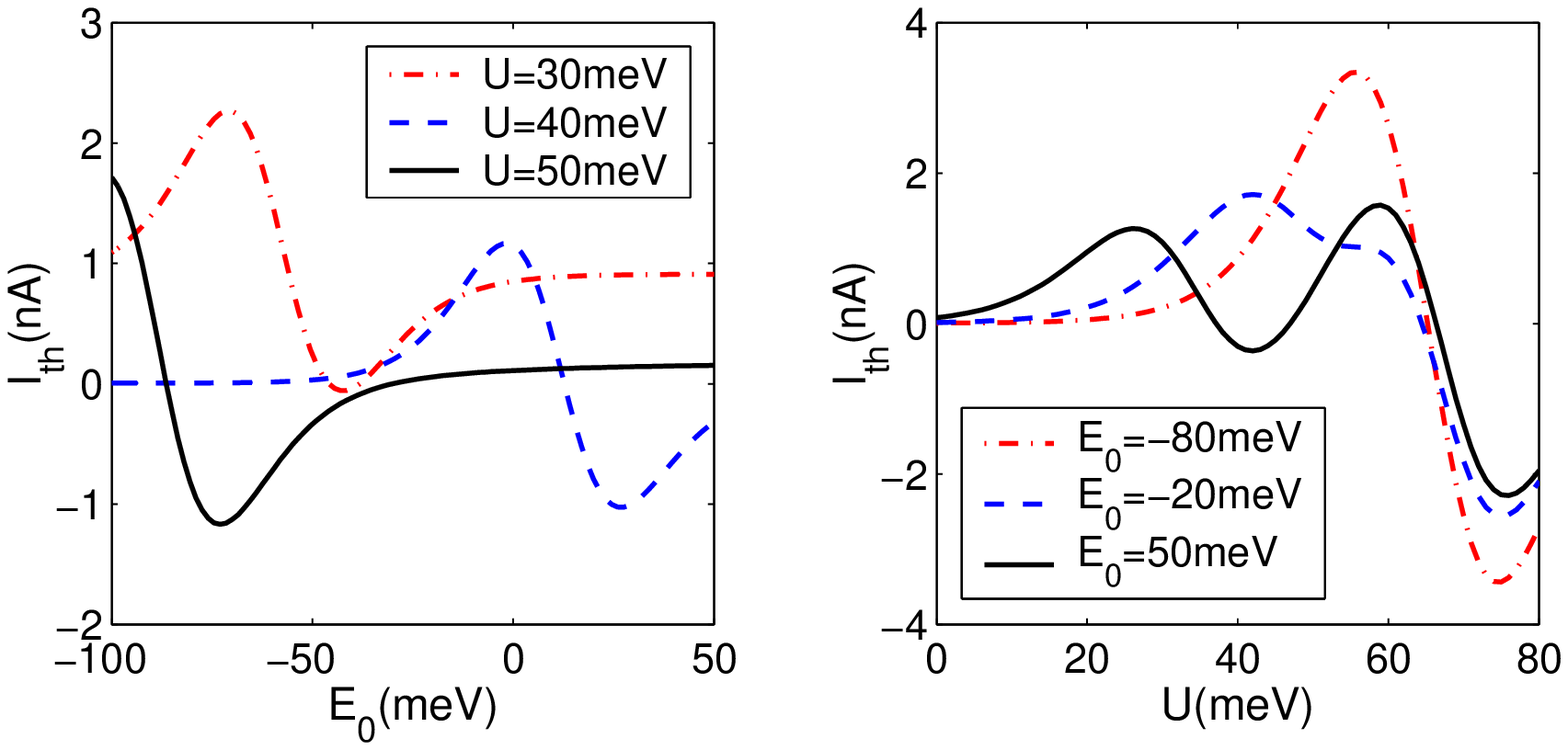} 
\caption{(Color online)  Top panel: Thermally excited current $ I_{th} $ versus $ E_0 $ and $ U. $ The surface is plotted assuming $ kT_R = 6meV,\ \Delta T = 0.8 T_R,\ \Gamma = 2.5meV,\ \Lambda = 7.5meV,\   \hbar\Omega = 10meV.$ Bottom panels: Cross-sections of the surface shown on the top at several fixed values of $ U  $ (left) and at several fixed values of $ E_0 $ (right).
}
 \label{rateI}
\end{center}\end{figure}

By simultaneously varying $ E_0 $ and $ \Lambda $ one may put either $ \tilde E_0 $ or $ \tilde E_0 + \tilde U $ inside this window thus providing favorable conditions for thermally excited current to flow. In this way, one can manipulate both magnitude and direction of $ I_{th}. $ Also,  the surface presented in the Fig. 4 is built assuming that leads temperatures  are rather high, and the thermal energies $kT_{L,R} $ and $ k \Delta T $  accept values comparable to $ \hbar\Omega. $ Under these conditions, signatures of metastable transport channels may appear in $ I_{th}(E_0,\Lambda )$ along with (or, even, instead of) those of original channels existing in a molecular junction with a rigid (nonvibrating) bridge. This may happen because the suppression of metastable states contributions occurring at low leads temperatures \cite{18,26} becomes less effective when $ T_{L,R} $ and $ \Delta T $ are moderately high.
   		   In principle, one may similarly control $ I_{th} $ by varying $ E_0 $ and $ U $ at a certain fixed value of $ \Lambda. $ Again, $ I_{th} $ will flow through the system when the renormalized bridge level energy $ \tilde E_0 $  (or $ \tilde E_0 + \tilde U)$ is moved into the conduction window. Also, one may expect phonon sidebands signatures to appear when $ kT_R,\ k\Delta T $ an $ \Lambda $ accept values of the order of $ \hbar\Omega. $ As known, polaron formation on the bridge may cause the reversal of the renormalized charging energy sign, so that electron repulsion is replaced by effective attraction. The crossover occurs when $ U = 2\Lambda^2/\hbar\Omega\ (\tilde U = 0). $  One observes that two ridges on the surface $I_{th} (E_0,U) $ displayed in 
			the Fig. 5 approach each other in the vicinity of the crossover, and they merge into a single peak at a certain value of the bridge level energy.

\section{iv. conclusion}

In this work, we theoretically studied the steady thermoelectric transport in a single-molecule junction with a vibrating bridge. In these studies, we employed a simple model for the bridge simulating it by a single spin degenerated energy level coupled to a sole vibrational mode. We concentrated on analysis of diverse properties of thermally excited charge current which appears when a temperature gradient is applied across the system. This quantity is especially interesting for it is available for direct measuring in experiments on nanoscale systems. 

The present studies were carried out taking into account both Coulomb repulsion between electrons on the bridge and their  interactions with the vibrational mode. The adopted computational method is based on nonequilibrium Green's functions formalism. To simplify calculations, only terms of lowest order in $ \Lambda/\hbar\Omega $ are kept in the expressions for the relevant Green's functions. Also, rather simple approximations are used for electron Green's functions $ \tilde G^<(E) $ and $ \tilde G^>(E). $
We realize limitations of the computational scheme employed in the present work. However, we remark that approximations for the electron Green's functions used in this work bring reasonably good results for characteristics of electron transport through QDs and single-molecule junctions, as shown in Ref. \cite{42} and several other works. These approximations remain appropriate when the considered system is not too strongly coupled, so that $ \Gamma $ is smaller than $ U. $ 

The effect of vibrational mode on the thermally excited current flowing through a single-molecule junction is twofold. First, electron-vibron interactions bring renormalizations of the bridge energy level (polaronic shift) and of charging energy $ U $ and effectively reduce the bridge coupling to the leads. Secondly, extra channels for electron transport associated with metastable states $ E_n,\ E_m $ may open up.  Signatures of these states may appear as extra peaks in the electron transmission and extra steps in current-voltage curves in biased junctions. 

At low temperatures, signatures of these extra peaks do not explicitly appear in $ I_{th}(E_0) $ lineshapes due to the "floating" condition of the phonon bands which was  first discussed as applied to studies of linear electric transport through molecules \cite{18}. Thus the electron-vibron interactions mostly affect $ I_{th} $ magnitude and direction through the polaron shift of $ E_0 $ and renormalization of $ U. $ These renormalization may move $ \tilde E_0 $ and $ \tilde E_0 + \tilde U $ inside/outside the "conduction window" occurring about the leads chemical potential when the temperature gradient is applied across the system. Thus electron-phonon interactions may affect both magnitude and direction of the thermally induced current at fixed $ E_0 $ value. The $ I_{th} $ dependence  of $ \Delta T $ appears to be nonlinear and nonmonotonic one. The magnitude and direction of thermally induced current are very sensitive to 
the values of relevant energies $ E_0,\ U ,\ \Lambda $ and $ k\Delta T. $ At moderately high temperatures of the leads when $ k\Delta T $ accepts values comparable to the energy of the vibrational phonon $ \hbar\Omega, $ signatures of metastable states may appear in $ I_{th}(E_o,\Lambda) $ bringing additional features into this surface shape. 

It was established in previous studies that Coulomb interactions between electrons on the bridge strongly influence thermoelectric properties of molecular junctions \cite{7,12,21,35,41,42}. Here, we analyzed the combined effect of Coulomb repulsion between electrons on the bridge and electron-phonon interactions on the thermally induced current through the system. The most interesting effects may appear when strong electron-phonon interactions cause the change of sign of the renormalized charging energy $ \tilde U. $ It was already shown that one may expect a significant enhancement of the linear response thermopower in a strongly coupled system described by an Anderson model with the negative charging energy \cite{43}. Therefore, it would be interesting to generalize the present analysis to strongly coupled systems which could be moved from the Coulomb blockade regime to Kondo regime and consider the effects which may occur when the renormalized charging energy $ \tilde U $ becomes negative due to electron-phonon interactions.

Finally, in this work we used a simple Anderson-Holstein model for a molecular junction, and a rather simple computational method was applied to derive approximations for the relevant Green's functions. Nevertheless, the presented analysis captures some essential physics associated with thermoelectric transport through vibrating molecules. We believe that the reported results may be useful for better understanding of nonlinear Seebeck effect in nanoscale systems.
\vspace{2mm}

 {\bf Acknowledgments:}
The author  thank  G. M. Zimbovsky for help with the manuscript. This work was supported  by  NSF-DMR-PREM 1523463.

\end{document}